\documentclass[superscriptaddress,floatfix,twocolumn,showpacs,preprintnumbers,amsmath,amssymb,prl]{revtex4}

\usepackage{graphicx}
\usepackage{dcolumn}
\usepackage{bm}

\begin{document}

\preprint{}

\title{Thermodynamics of the Up-Up-Down Phase of the $\bm{S = \frac{1}{2}}$
Triangular-Lattice Antiferromagnet Cs$_2$CuBr$_4$}

\author{H. Tsujii}
\affiliation{Department of Physics, University of Florida, P.O.\ Box 118440,
Gainesville, Florida 32611-8440, USA}
\affiliation{Department of Physics, Kanazawa University, Kakuma-machi,
Kanazawa 920-1192, Japan}
\author{C. R. Rotundu}
\altaffiliation[Present address: ]{Department of Physics, University of Maryland, College Park, MD 20742-4111,
USA.}
\affiliation{Department of Physics, University of Florida, P.O.\ Box 118440, Gainesville, Florida
32611-8440, USA}
\author{T. Ono}
\author{H. Tanaka}
\affiliation{Department of Physics, Tokyo Institute of Technology, Meguro-ku,
Tokyo 152-8551, Japan}
\author{B. Andraka}
\author{K. Ingersent}
\author{Y. Takano}
\affiliation{Department of Physics, University of Florida, P.O.\ Box 118440,
Gainesville, Florida 32611-8440, USA}

\date{\today}

\begin{abstract}
Specific heat and the magnetocaloric effect are used to probe the
field-induced up-up-down phase of Cs$_2$CuBr$_4$, a quasi-two-dimensional
spin-$\frac{1}{2}$ triangular antiferromagnet with near-maximal frustration.
The shape of the magnetic phase diagram shows that the phase
is stabilized by quantum fluctuations, not by thermal fluctuations as
in the corresponding phase of classical spins. The magnon gaps
determined from the specific heat are considerably larger than those
expected for a Heisenberg antiferromagnet, probably due to the presence of a
small Dzyaloshinskii-Moriya interaction.
\end{abstract}


\maketitle

The interplay between geometric frustration and quantum fluctuations in
small-spin antiferromagnets provides fertile ground for observation of new
phenomena. The prime example is a spin $S=\frac{1}{2}$ antiferromagnet on a
triangular lattice, which has been intensively studied since Anderson's
conjecture of a resonating-valence-bond ground state \cite{Anderson}.
The zero-field ground state of the nearest-neighbor Heisenberg model has been
shown to be weakly ordered with a 120$^{\circ}$ spin arrangement \cite{Bernu}.
However, the magnons suffer from unusual two-particle decay processes
and display significantly renormalized dispersion \cite{Chernyshev} with
roton-like minima at the zone boundaries \cite{StarykhPRB,Zheng06}.
Experimentally, observation of unusual dynamics in the spin-$\frac{1}{2}$
triangular antiferromagnet Cs$_2$CuCl$_4$ \cite{Coldea01} has led to proposals
of nearby spin-liquid states \cite{Isakov,Alicea}.

The frustration-fluctuation interplay in $S=\frac{1}{2}$ triangular antiferromagnets also manifests itself in a
magnetization plateau at 1/3 of the saturation value in both Heisenberg and XY nearest-neighbor models
\cite{Chubukov}. Any magnetization plateau must arise from an energy gap in the low-lying magnetic excitations.
Since such a gap is a consequence of the ground state maintaining the continuous rotational symmetry of the
Hamiltonian, a magnetization plateau indicates that the ground state is a spin liquid, a collection of spin
multimers, or an ordered state that is collinear with the magnetic field. Moreover, the ground state must be
commensurate with the underlying crystal lattice, unless it is a spin liquid \cite{Oshikawa00}. In
spin-$\frac{1}{2}$ Heisenberg and XY antiferromagnets on a triangular lattice, the plateau arises from a
collinear up-up-down (uud) phase \cite{Chubukov}, in which up spins parallel to the magnetic field form a
honeycomb sublattice and the down spins form a triangular sublattice comprising the centers of the hexagonal
honeycomb cells.

Among the known spin-$\frac{1}{2}$ triangular-lattice antiferromagnets, Cs$_2$CuBr$_4$ is the only one
exhibiting a magnetization plateau indicative of the uud phase \cite{Ono03,Ono04,Ono05JPSJsuppl}. The compound
has an orthorhombic crystal structure with space group $Pnma$ \cite{Morosin}. The magnetic Cu$^{2+}$ ions are
located within distorted CuBr$_4^{2-}$ tetrahedra, which form a triangular lattice in the $bc$ plane. At the
magnetization plateau with $\bm{H}\,||\,\bm{c}$, the $b$ component of the order vector detected by neutrons
agrees within the experimental uncertainty with the wave number $k_0=2/3$ of the uud phase \cite{Ono04}. The
$^{133}$Cs NMR spectra for $\bm{H}\,||\,\bm{b}$ provide further evidence for this phase \cite{Fujii}.

In Cs$_2$CuBr$_4$, the nearest-neighbor Cu$^{2+}$ exchange $J_1$ along $b$ is greater than $J_2$ along other
principal directions in the $bc$ plane. The ratio $J_2/J_1$ is 0.74, according to a comparison
\cite{Ono05JPSJsuppl} of the wave number $k_0$ of the incommensurate, cycloidal ordered structure at zero field
\cite{Ono04} with results of linked-cluster expansions \cite{Zheng99}. Therefore, Cs$_2$CuBr$_4$ is much closer
to the maximally frustrated limit $J_2/J_1=1$ than is the extensively studied analog Cs$_2$CuCl$_4$
\cite{Coldea01,Coldea,Coldea02,Radu,Tokiwa}, for which $J_2/J_1 = 0.34$--0.37 \cite{Ono05JPSJsuppl,Coldea02}.
Numerical diagonalization of finite-size spin-$\frac{1}{2}$ Heisenberg systems predicts that the geometric
frustration is sufficient to stabilize the uud phase only in the range $0.7\lesssim J_2/J_1\lesssim 1.3$
\cite{Miyahara}, explaining its presence in Cs$_2$CuBr$_4$ and absence in the chloride. However, this prediction
is challenged by a renormalization-group calculation \cite{StarykhUnpbl} that finds the uud phase for
infinitesimally small $J_2$ \cite{Starykh}.

In this Letter, we report the unique thermodynamic properties of the uud phase
of Cs$_2$CuBr$_4$ based on magnetocaloric-effect and specific-heat measurements.
We examine in detail the phase diagram, which strongly differs from that of
classical spins, to uncover the role of quantum fluctuations in stabilizing
this phase. The specific heat reveals dramatic enhancement of the magnon gap,
which we attribute to the presence of a weak Dzyaloshinskii-Moriya (DM)
interaction \cite{DM,weakDM}. Recently, a novel spin liquid and a weak uud
order have been proposed to occur at 1/3 of the saturation magnetization for
$J_2/J_1<1$ \cite{Alicea07}. The specific-heat data suggest that the uud phase
of Cs$_2$CuBr$_4$ is far from such exotic states, at least for the field
orientation of the present study. Preliminary results were presented in
\cite{LT24}.

The experiment was performed in magnetic fields applied along the $c$ axis.
The sample-growth method \cite{Ono03} and the calorimeter \cite{Tsujii03} have
been previously described.

\begin{figure}[btp]
\begin{center}\leavevmode
\includegraphics[width=0.6\linewidth,,angle=-90]{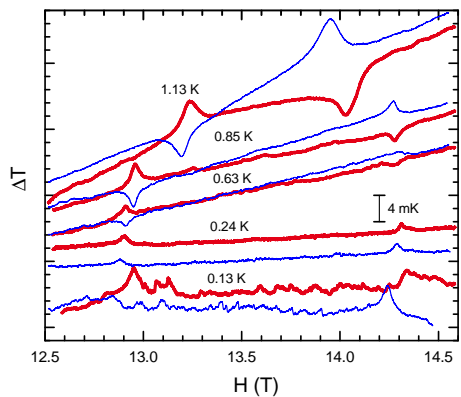}
\caption{(Color online) Magnetocaloric effect of Cs$_2$CuBr$_4$ in fields along the $c$ axis: temperature
difference $\Delta T$ between sample and thermal reservoir during 0.2-T/min upward and downward field sweeps
(thick and thin lines, respectively). Traces at different temperatures are offset for clarity. The overall
separation between up- and down-sweep traces at the lowest temperatures is an uninteresting nuclear-spin
effect.}\label{fig1}\end{center}\end{figure}

Figure \ref{fig1} shows the magnetocaloric-effect results, where we swept the magnetic field between 12.5\,T and
14.6\,T at a rate of 0.2\,T/min, while continuously measuring the temperature difference between the sample and
the thermal reservoir. At temperatures $T\le 1.17$\,K the transitions between the uud and other phases, which
are incommensurate states according to neutron diffraction \cite{Ono04} and $^{133}$Cs NMR \cite{Fujii}, appear
as peaks and dips in the temperature traces. At 0.24\,K and 0.13\,K, the two lowest temperatures of the
experiment, the transitions become clearly hysteretic: features indicating transitions during up sweeps are
shifted relative to those during down sweeps; in addition, all transitions appear as peaks irrespective of the
field-sweep direction, indicating heating due to irreversibility. These two signatures of hysteresis
unambiguously indicate that the transitions between the uud and incommensurate phases are first-order, as
previously suggested by magnetization and elastic-neutron data \cite{Ono05JPSJsuppl}. The absence of detectable
hysteresis for $T\ge 0.63$\,K suggests that at these higher temperatures the nucleation rate of a new phase at
each transition field exceeds the field-sweep rate.

\begin{figure}[btp]
\begin{center}\leavevmode
\includegraphics[width=0.6\linewidth,angle=-90]{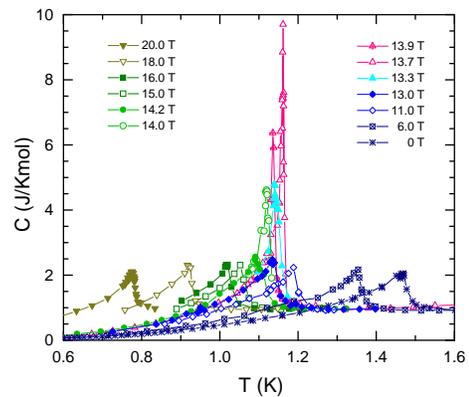}
\caption{(Color online) Magnetic specific heat of Cs$_2$CuBr$_4$ in zero magnetic field and in fields along the
$c$ axis. A small phonon contribution, $7.94 T^3$\,mJ/K\,mol, has been subtracted by scaling the specific heat
of Cs$_2$ZnBr$_4$ \cite{Ono05JPSJsuppl}. The lines are guides to the eye.}\label{fig2}\end{center}\end{figure}

The transition from the high-temperature, paramagnetic phase to the antiferromagnetically ordered phases appears
as a peak in the specific heat at all magnetic fields up to 20\,T, the highest field of the present study. As
shown in Fig.\ \ref{fig2}, the peak height at the transition is nearly the same throughout the incommensurate
phases, whereas it is larger and strongly field-dependent in the uud phase, reaching a maximum at 13.7\,T.
Although relaxation calorimetry is generally poor at distinguishing a sharp specific-heat peak from a latent
heat, it is quite likely that, like the incommensurate-uud transition, the paramagnetic-uud transition is
first-order. The robustness of this transition argues against a proximity of this system to the exotic phases
recently proposed \cite{Alicea07}.

\begin{figure}[btp]
\begin{center}\leavevmode
\includegraphics[width=0.65\linewidth]{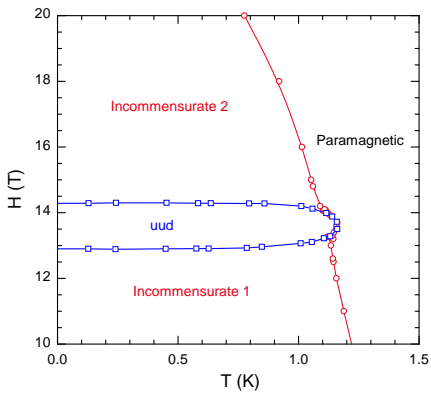}
\caption{(Color online) Phase diagram of Cs$_2$CuBr$_4$ in magnetic fields along the $c$ axis, as deduced from
magnetocaloric-effect (squares) and specific-heat (circles) measurements. Lines indicating the phase boundaries
are guides to the eye.}\label{fig3}\end{center}\end{figure}

From the positions of the specific-heat peaks and the sharp features in the
magnetocaloric-effect temperature traces, we obtain the magnetic phase diagram
shown in Fig.\ \ref{fig3}. Below 0.7\,K, the phase boundaries of the uud phase
are nearly horizontal, indicating (via the magnetic Clausius-Clapeyron relation)
a very small entropy difference between this phase and the incommensurate
phases. The critical fields extrapolated to $T=0$ are $H_{c1}=12.9$\,T and
$H_{c2}=14.3$\,T, slightly lower than the values 13.1\,T and 14.4\,T
obtained from the magnetization curve \cite{Ono05JPSJsuppl}.

Above 0.7\,K, the width of the uud phase decreases slightly with increasing temperature, indicating that this
phase has a smaller entropy than the incommensurate phases. The uud phase is thus seen to owe its existence to
an energy-lowering mechanism rather than to thermal fluctuations, which would decrease the free energy by
raising the entropy. That the reduced energy of the uud state in comparison with the incommensurate phases more
than compensates for its slightly lower entropy is confirmed by a bulge of the uud phase into the paramagnetic
phase. According to spin-wave theory \cite{Chubukov}, this energy lowering is due to quantum fluctuations. The
shapes of the observed phase boundaries are in marked contrast to those of the uud phase of classical spins
\cite{Kawamura85B,Korshunov}. Being a single point in the $T=0$ phase diagram, the classical uud phase becomes
stable only at nonzero temperatures as thermal fluctuations raise its entropy relative to that in either
adjacent ordered phase. As a result, the field width of the phase expands with increasing temperature, as
observed for instance in RbFe(MoO$_4$)$_2$ \cite{Svistov,Smirnov}.

The lowering of the energy of the uud phase explains why the transitions at $H_{c1}$ and $H_{c2}$ are
first-order. For classical spins, the ground state and, with it, the magnetization evolve continuously with
magnetic field, the uud state being a ground state only at one field. As the energy of this state is
preferentially lowered by quantum fluctuations, leaving behind some states over a range of magnetization values,
these states lose their ability to be
a ground state. Consequently, the wave function and the magnetization of the ground state change
discontinuously at the critical fields.

The zero-temperature width of the uud phase, $H_{c2}-H_{c1}$, is directly related to magnon gaps.
In particular, if the spin Hamiltonian commutes with the total spin, the gaps are
$\textsl{g}\mu_{\mathrm B}(H-H_{c1})$ and $\textsl{g}\mu_{\mathrm B}(H_{c2}-H)$
for the $S_z=-1$ and $S_z=+1$ magnons, respectively. Here $\textsl{g}$ is the \textrm{g} factor and
$\mu_{\rm B}$ the Bohr magneton.

\begin{figure}[btp]
\begin{center}\leavevmode
\includegraphics[width=0.7\linewidth]{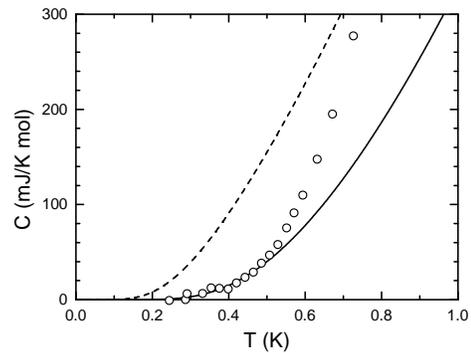}
\caption{Magnetic specific heat of Cs$_2$CuBr$_4$ in a 13.7-tesla field along the $c$ axis. The lines are
calculations for gapped magnons, as discussed in the text.}\label{fig4}\end{center}\end{figure}

The presence of the gaps is evident in the low-temperature specific heat at 13.7\,T (roughly the mid-point of
the uud phase) shown in Fig.\ \ref{fig4}. Here, the nuclear-spin contribution, 21.2$(H/T)^2$\,$\mu$J/K\,mol with
$H$ in tesla and $T$ in kelvin \cite{Bennett}, has been subtracted along with the insignificant phonon
contribution of 7.94~$T^3$~mJ/K\,mol. In the nuclear contribution, we have ignored hyperfine interactions and
quadrupole interactions. This approximation is justified, since it gives a nuclear specific heat for
Cs$_2$CuCl$_4$ that agrees to within 23\% with the experimental data \cite{Radu}. We expect the approximation to
be considerably better for Cs$_2$CuBr$_4$, where the nuclear contribution is dominated by $^{79}$Br and
$^{81}$Br having quadrupole moments an order of magnitude smaller than those of $^{35}$Cl and $^{37}$Cl
\cite{Quadru}.

\begin{figure}[btp]
\begin{center}\leavevmode
\includegraphics*[viewport=0 320 800 720, width=0.95\linewidth]{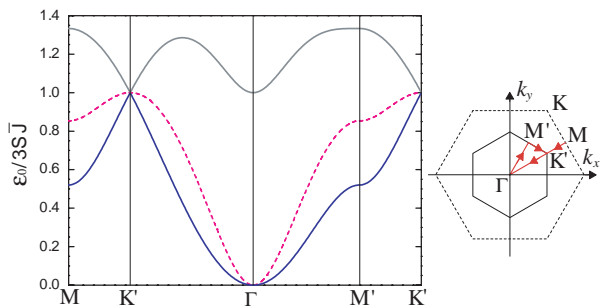}
\caption{(Color online) Dispersion $\epsilon_0(k)$ of the $S_z=-1$ (solid lines)
and $S_z=+1$ (broken line) classical magnon modes in the uud state of a
Heisenberg antiferromagnet on a triangular lattice, along the high-symmetry
paths indicated with arrows in the reciprocal space at right.
In the reciprocal space, the solid line outlines the first Brillouin zone
of the three triangular sublattices, and the broken line that of the
full lattice.}\label{fig5}\end{center}\end{figure}

The magnon dispersion of the uud phase is known for the classical Heisenberg
model \cite{Chubukov}. Anisotropy may be accounted for by substituting the
average exchange $\bar{J}$=$(J_1+2J_2)/3$ for the isotropic $J$. As shown in
Fig.\ \ref{fig5}, the two low-energy modes have significantly different
dispersions, reflecting the different symmetries of the coplanar phases that
occur below and above the field region of the uud phase. Near the zone center,
the dispersion is $\epsilon_{0-}(\bm{k})\simeq(3/4)S\bar{J}k^2$ for the lower
of two $S_z=-1$ modes and $\epsilon_{0+}(\bm{k})\simeq(9/4)S\bar{J}k^2$ for
the $S_z$=+1 mode.
These classical magnons are gapless, consistent with the collapse of the uud
phase to a single point in the phase diagram at $T=0$.
For $S=\frac{1}{2}$, however, quantum fluctuations give rise to gaps at the
zone center \cite{Chubukov}.

The $k$ dependence of the magnon dispersion is not known for $S=\frac{1}{2}$, but we expect
$\epsilon_{\pm}(\bm{k})=\epsilon_{0\pm}(\bm{k})+\Delta_{\pm}$ to be a good approximation for the low-energy
$S_z=\pm 1$ modes, where $\Delta_{\pm}$ are the gaps. We ignore the higher-energy $S_z=-1$ mode, since its
contribution to the specific heat is negligible.

To quantitatively compare the data with a gapped-magnon behavior, we need to know the exchange couplings. These
can be determined from $J_2/J_1$ and from the measured $\textsl{g}H_s\simeq 63$\,T, which holds for all three
principal field directions despite small variations in $\textsl{g}$ and in the saturation field $H_s$
\cite{Ono03}, combined with the theoretical result $H_s=J_1(2+J_2/J_1)^2/(2\textsl{g}\mu_{\mathrm B})$, which is
exact when terms other than $J_1$ and $J_2$ are negligible in the spin Hamiltonian. The results are
$J_1=11.3$\,K and $J_2=8.3$\,K \cite{error}, yielding a value $\bar{J}=9.3$\,K for substitution into
$\epsilon_{0\pm}(\bm{k})$.

Finally, the magnon specific heat is given by
\begin{equation}
C(T)=\frac{R}{3A_k}\sum_{s=\pm}\int d^2\bm{k}
\left[\frac{\beta\epsilon_s(\bm{k})}{e^{\beta\epsilon_s(\bm{k})/2}-
e^{-\beta\epsilon_s(\bm{k})/2}}\right]^2
\end{equation}
at low temperatures, where interactions between magnons can be ignored. Here, $R$ is the gas constant,
$\beta=1/k_{\mathrm B}T$, and the integral is performed numerically over the first Brillouin zone $A_k$ of the
sublattices. As shown by the broken line in Fig.\ \ref{fig4}, $\Delta_{-}=\textsl{g}\mu_{\mathrm B}(H-H_{c1})$
and $\Delta_{+}=\textsl{g}\mu_{\mathrm B}(H_{c2}-H)$, with $\textsl{g}=2.24$ from ESR \cite{Nojiri} and $H_{c1}$
and $H_{c2}$ taken from the present work, give too large a specific heat in comparison with the data.
Surprisingly, the best fit requires gaps that are $2.1\pm 0.1$ times these values, as shown by the solid line.
According to spin-wave theory \cite{Chubukov}, the linear field dependence of the gaps breaks down in
spin-$\frac{1}{2}$ XY antiferromagnets, whose Hamiltonian does not commute with the total spin, giving way to
enhanced gaps proportional to $|H/H_{c1,2}-1|^{1/2}$. This prediction suggests that a weak DM interaction, which
also introduces an easy-plane anisotropy (albeit different from an XY type), is responsible for the large gaps
found in the specific heat. It is intriguing that, at the same time, this anisotropy destroys the uud phase when
$\bm{H}\,||\,\bm{a}$ \cite{Ono05JPSJsuppl}.

In summary, we have studied the thermodynamics of the uud phase of the spin-$\frac{1}{2}$ triangular-lattice
antiferromagnet Cs$_2$CuBr$_4$. The shape of the phase diagram implies that this phase is stabilized primarily
by quantum fluctuations. The transitions to the phase from the incommensurate phases and quite possibly from the
high-temperature, paramagnetic phase are first-order as a result of quantum fluctuations, in contrast to the
second-order transitions from the paramagnetic phase to the incommensurate phases. The gaps for the two
low-energy magnon modes are considerably larger than expected from the field width of the uud phase, suggesting
gap enhancement by the Dzyaloshinskii-Moriya interaction.

We thank L.\ Balents, A.\ L.\ Chernyshev, A.\ V.\ Chubukov, Y.\ Fujii,
D.\ L.\ Maslov, S.\ Miyahara, J.\ L.\ Musfeldt, M.\ Oshikawa, O.\ A.\ Starykh,
M.\ Takigawa, and N.\ Todoroki for useful discussions, and T.\ P.\ Murphy
and E.\ C.\ Palm for technical assistance. This work was supported by the NSF
through DMR-9802050 (YT) and DMR-0312939 (KI), and by the DOE under Grant
No.\ DE-FG02-99ER45748 (BA). A portion of the work was carried out at the
National High Magnetic Field Laboratory, which is supported by NSF Cooperative
Agreement No. DMR-0084173 and by the State of Florida.


\begin{thebibliography}{99}
\bibitem{weakDM}
The strength $D$ of the DM interaction in Cs$_2$CuCl$_4$ is $0.054J_1$ \protect\cite{Coldea02}. Its strength in
Cs$_2$CuBr$_4$ has not been determined.
\bibitem{error}
The values of $J_1$ and $J_2$ given in \protect\cite{Ono03}
and \protect\cite{Ono05JPSJsuppl} are incorrect due to a numerical error.
\end{thebibliography}
\end{document}